\documentclass[epj]{svjour}
\usepackage{graphicx}
\usepackage{siunitx}
\usepackage{mhchem}
\usepackage{xspace}
\usepackage{booktabs}
\usepackage[flushleft]{threeparttable}

\usepackage[draft]{fixme}
\newcommand{\unsim}{\mathord{\sim}}
\newcommand{\ECM}{E_{\mathrm{CM}}}
\newcommand{\Be}{\ce{^{8}Be}\xspace}
\newcommand{\Boron}{\ce{^{11}B}\xspace}
\newcommand{\BoronT}{\ce{^{12}B}\xspace}
\newcommand{\Carbon}{\ce{^{12}C}\xspace}
\newcommand{\cm}{\mathrm{CM}}

\newcommand{\mref}{ref. }
\newcommand{\mrefs}{refs. }
\newcommand{\mfig}{fig.}
\newcommand{\fref}[1]{\mfig~\ref{#1}}
\newcommand{\meqref}[1]{eq.~\eqref{#1}}

\newcommand{\tref}[1]{table~\ref{#1}}
\newcommand{\etal}{\textit{et al.}\xspace}
\newcommand{\mb}{\milli \barn}
\newcommand{\ub}{\micro \barn}
\newcommand{\Glab}{\Gamma_{\mathrm{lab}}}

\makeatletter
\newcommand{\lambdabar}{{\mathchoice
    {\smash@bar\textfont\displaystyle{0.25}{1.2}\lambda}
    {\smash@bar\textfont\textstyle{0.25}{1.2}\lambda}
    {\smash@bar\scriptfont\scriptstyle{0.25}{1.2}\lambda}
    {\smash@bar\scriptscriptfont\scriptscriptstyle{0.25}{1.2}\lambda}
  }}
\newcommand{\smash@bar}[4]{%
  \smash{\rlap{\raisebox{-#3\fontdimen5#10}{$\m@th#2\mkern#4mu\mathchar'26$}}}%
}
\makeatother

\begin{document}
\title{
  The partial widths of the 16.1 MeV 2\textsuperscript{+} resonance in \Carbon
}

\author{
  Michael Munch\inst{1}
  and Hans Otto Uldall Fynbo\inst{1}
}                     
\authorrunning{M. Munch and H. O. U. Fynbo}
\offprints{Hans Fynbo}          
\institute{Department of Physics and Astronomy, Aarhus University, Ny Munkegade 120, 8000
  Aarhus C, Denmark}
\date{Received: date / Revised version: date}
%
\abstract{
  The \SI{16.1}{\MeV} $2^{+}$ resonance in \Carbon situated slightly
  above the proton threshold can decay by proton-, $\alpha$-, and
  $\gamma$ emission. The partial width for proton emission cannot be
  directly measured due to the low proton energy and the small
  branching ratio. Instead it must be indirectly derived from other
  observables. However, due to several inconsistent data the derived
  partial width varies by almost a factor 2 dependent on the data
  used. Here we trace the majority of this inconsistency to different
  measurements of the $(p,\alpha)$ cross sections.  We have remeasured
  this cross section using modern large area silicon strip detectors
  allowing to measure all final state particles, which circumvents a
  normalization issue affecting some of the previous
  measurements. Based on this we determine $\Gamma_{p} =
  \SI{21.0(13)}{\eV}$. We discuss the implications for other
  observables related to the \SI{16.1}{\MeV} $2^{+}$ resonance
  and for isospin symmetry in the $A=12$ system.
  In addition, we conclude that the dataset currently used for the NACRE and NACRE
  II evaluation of the $\Boron(p,3\alpha)$ reaction should be scaled by a factor of $2/3$. This
  impacts the reaction rate accordingly. 
} 
\maketitle
\section{Introduction}
\label{intro}

\begin{table*}[!h]
  \centering
  \caption{Prior measurements of the $(p,\alpha)$ channel. }
  \label{tab:prev}
  \begin{threeparttable}
  \begin{tabular*}{\textwidth}{l@{\extracolsep{\fill}}ccccc}
    \toprule
    Measurement                                 & $\sigma_{\alpha,0}$ [\si{\mb}] & $\sigma_{\alpha,1}$ [\si{\mb}] & $\sigma_{\alpha}$ [\si{\mb}] & $\sigma_{\alpha,1}/\sigma_{\alpha,0}$ & $\Gamma$ [\si{\keV}]    \\
    \midrule
    Huus \etal \cite{Huus1953}   & &  &                    &                   &   $\unsim 5$                 \\
    Beckman \etal \cite{Beckman1953}   & $\num{0.2} \pm 30\%$\tnote{1}   & $\num{10} \pm 30\%$\tnote{1}    &                    &                   &                    \\
    Segel \etal \cite{Segel1961}       &                      &                      &                    & \num{22(3)}       &                    \\
    Anderson \etal \cite{Anderson1974} &                      &                      & \num{41(3)}\tnote{2}     &                   & \num{6.7}          \\
    Davidson \etal \cite{Davidson1979} &                      &                      & \num{54(6)}        &                   & $5.2\substack{+0.5 \\ -0.3}$       \\
    Becker \etal \cite{Becker1987}     & $\num{2.12} \pm 5\%$   & $\num{69.6} \pm 5\%$\tnote{3}   &                    &  \num{33(2)}                 & \num{5.3(2)}       \\
    Laursen \etal \cite{Laursen2016}   &                      &                      &                    & \num{19.6(19)}       &                    \\
    \bottomrule
  \end{tabular*}
  \begin{tablenotes}
  \item[1] The authors note that the $\alpha$ particles were ``barely detectable'' \cite{Beckman1953}. This result will be disregarded.
  \item[2] Assuming infinite target thickness and using the combined $\Gamma$ of
    \mref \cite{Anderson1974,Becker1987} this should be rescaled from \SI{38.5(32)}{\mb}.
  \item[3] The authors note that their model did \emph{not} reproduce the $\alpha_{1}$ data.
  \end{tablenotes} 
  \end{threeparttable}
\end{table*}

\begin{figure}[b]
  \includegraphics[width=\columnwidth]{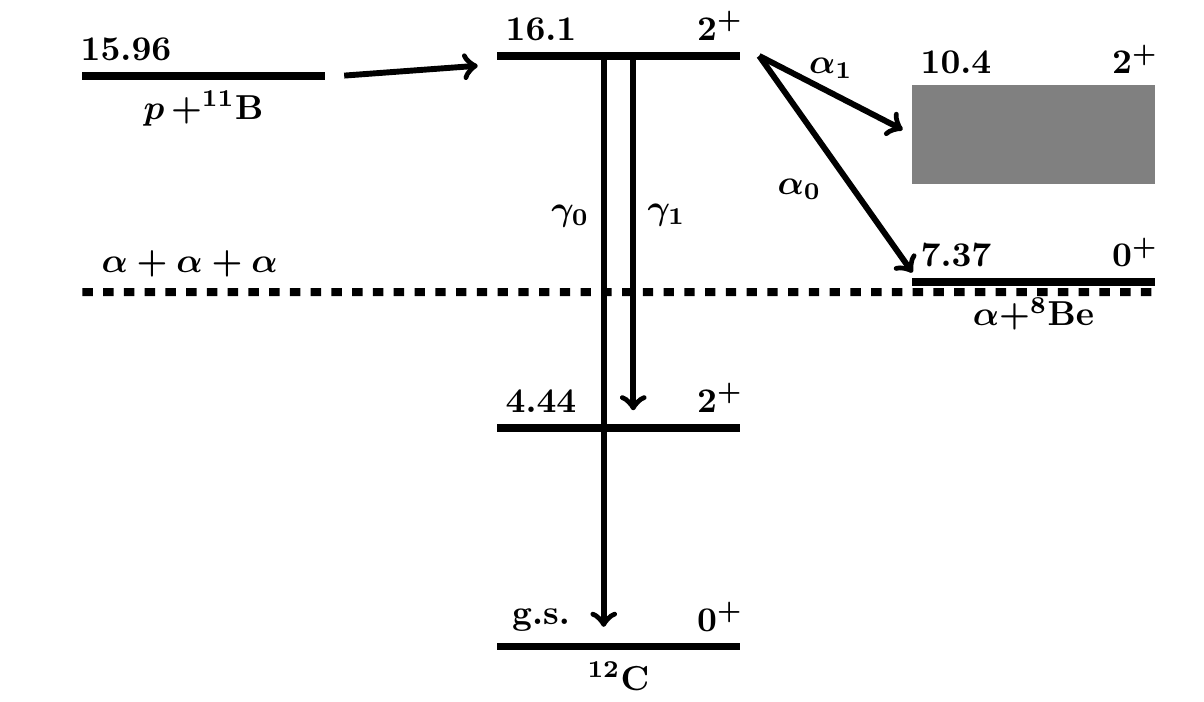}
  \caption{Illustration of the reaction scheme. The \SI{16.11}{\MeV} $2^{+}$ state \Carbon is
    populated with the $p+\ce{^{11}B}$ reaction. The state can either decay via $\gamma$, $\alpha$
    or proton emission. Energies and spin assignments are taken from
    \mref \cite{Kelley2017}. Energies are in MeV. 
  }
  \label{fig:scheme}       
\end{figure}

Situated just above the proton threshold the \SI{16.1}{\MeV} $2^{+}$
state in \Carbon has been the subject of numerous studies
\cite{Huus1953,Beckman1953,Segel1961,Anderson1974,Davidson1979,Becker1987,Laursen2016,Cecil1992,Adelberger1977,Friebel1978,%
  Stave2011,Laursen2016b,He2016,Craig1956}
with the most recent compilation published in \mref \cite{Kelley2017}
and a detailed review of the decay properties of the \SI{16.1}{\MeV}
state given in \mref \cite{Laursen2016b}. The state has primarily been
investigated with the $p + \ce{^{11}B}$ reaction and it is known to
decay via proton, $\alpha$ particle and $\gamma$ ray emission as
illustrated in \fref{fig:scheme}.

In one of the first applications of the concept of isospin, the narrow 
width of only roughly \SI{5}{\keV} of this state situated higher than 
\SI{8}{MeV} in the 3$\alpha$ continuum was explained by Oppenheimer 
and Serber to be due to its $T=1$ nature \cite{Oppenheimer1938}. Hence, 
its dominating $\alpha$-decay mode is only possible due to admixtures 
of $T=0$ in the state.  

In the narrow resonance limit the measured cross sections,
$\sigma_{px}$, can be related to the partial widths, $\Gamma_{x}$, of
the resonance
\begin{equation}
  \label{eq:width}
  \sigma_{px} = 4\pi\lambdabar^{2}\omega \frac{\Gamma_{p}\Gamma_{x}}{\Gamma^{2}},
\end{equation}
where $\omega = \frac{2J+1}{(2j_{0}+1)(2j_{1} + 1)}$ with $J$ the resonance spin and $j_{i}$ the spin
of the beam and target. $\lambdabar = \hbar/E$ is the reduced de Broglie
wavelength with the center of mass energy, $E$.
$\Gamma_{p}$ has a key role in this relation, but due to the low proton energy and the fact that
$\Gamma_{p}/\Gamma \ll 1$, it is not feasible to measure it directly. Instead, $\Gamma_{p}$ must be inferred from
measurements of both $\Gamma_{x}$ and $\sigma_{px}$. This extraction
was performed in both \mrefs \cite{Laursen2016b,Kelley2017},
however, the resulting proton widths differ by almost a factor of 2.

The decay properties of $T=1$ isobaric analog states in the $A=12$
system was analysed by Monahan \etal \cite{Monahan1971}. This analysis
was based on a comparison of the proton widths in $\Carbon$ with the
neutron spectroscopic factors in $\BoronT$ deduced from the
$\Boron(d,p)\BoronT$ reaction. Good agreement was found for most states
with the notable exception of the \SI{16.1}{Mev} state. The
discrepancy was traced to a too large value for the proton
width. Recently the $\Boron(d,p)\BoronT$ reaction was remeasured with a
new method which confirmed the spectroscopic factors deduced
previously within 25\% \cite{Lee2010}. The proton width recommended by
\mref \cite{Kelley2017}  results in good agreement with this spectroscopic
factor, while that of \mref \cite{Laursen2016b} does not.  
In the following we will summarize the results of previous
measurements and attempt to clarify the situation. 
 
$\gamma$ ray emission predominantly occur to the ground state (GS), $\gamma_{0}$, and the first excited
state, $\gamma_{1}$, in \Carbon. The cross sections for the $(p,\gamma)$ reactions were most recently
measured by He \etal using a thin target for the first time \cite{He2016}.
Here they confirmed the prior cross section measurements
\cite{Huus1953,Anderson1974,Cecil1992,Adelberger1977,Craig1956,Ajzenberg1952,Ajzenberg-Selove1990}
yielding a combined result of $\sigma_{p\gamma0} = \SI{5.1(5)}{\ub}$ and $\sigma_{p\gamma1} =
\SI{139(12)}{\ub}$. Thus we consider the values for these cross sections to be reliable. Complementary to these
measurements, Friebel \etal directly measured
$\Gamma_{\gamma0} = \SI{0.346(41)}{\eV}$ using inelastically scattered electrons
\cite{Friebel1978}, while Cecil \etal have measured the relative yield of $\gamma$
rays and charged particles; $\Gamma_{\gamma0}/\Gamma_{\alpha} = \num{6.7(3)E-5}$ and $\Gamma_{\gamma1}/\Gamma_{\alpha} =
\num{2.0(3)E-3}$ \cite{Cecil1992}.

The current understanding of the $\alpha$ particle decay mechanism is a sequential decay proceeding
either through the \Be GS, $\alpha_{0}$, or the first excited state, $\alpha_{1}$
\cite{Laursen2016}. The results of prior investigations of the $(p,\alpha)$ channel are listed in
\tref{tab:prev}. There are multiple consistent measurements for the resonance
width and combining the results from \mrefs \cite{Davidson1979,Becker1987} yields
\SI{5.28(18)}{\keV}. As these are extracted from a simple resonance scan we consider them
reliable. Two out of three measurements of the $\alpha_{1}/\alpha_{0}$ branching ratio are consistent and,
as the measurement by Laursen \etal was done with coincident detection of multiple
final state particles, we also consider the branching ratio reliable. The measured total cross
sections for $(p,\alpha)$ generally show poor agreement. However, considering the $(p,\alpha_{0})$ reaction
yields a distinct high energy peak we expect the $\sigma_{p,\alpha0}$ measurement by
Becker \etal to be accurate \cite{Becker1987}. 

By combining the various measurements for the $\alpha$- and $\gamma$ channels with \meqref{eq:width} and approximating $\Gamma\approx\Gamma_{\alpha}$ it is
possible to derive several independent values for $\Gamma_{p}$. These are listed in
\tref{tab:Gp}. Interestingly, the values seem to cluster into two groups, with the measurements
for $(p,\alpha)$ split across the groups.

\begin{table}[b]
  \centering
  \caption{Calculated values for $\Gamma_{p}$.
    The values are calculated using \meqref{eq:width} with the quantities listed in the left
    column. In all cases $\Gamma=\SI{5.28(18)}{\keV}$ was also used. The approximation $\Gamma_{\alpha}\approx\Gamma$ was
    applied. 
  }
  \label{tab:Gp}
  \begin{threeparttable}
    \begin{tabular*}{0.7\columnwidth}{l@{\extracolsep{\fill}}S}
      \toprule
      Method & {$\Gamma_{p}$ [eV]}\\
      \midrule
      $\sigma_{p\alpha}$ \cite{Anderson1974}                                               & 20(2) \\
      $\sigma_{p\alpha0}$ \cite{Becker1987} + $\Gamma_{\alpha1}/\Gamma_{a0}$ \cite{Segel1961,Laursen2016}   & 22(3) \\
      $\sigma_{p\alpha}$ \cite{Davidson1979}                                               & 26(3) \\
      $\sigma_{p\alpha}$ \cite{Becker1987}                                                 & 34(6) \\
      $\sigma_{p\gamma1}$ \cite{He2016} + $\Gamma_{\gamma1}/\Gamma_{a}$ \cite{Cecil1992}                    & 35(3) \\
      $\sigma_{p\gamma0}$ \cite{He2016} + $\Gamma_{\gamma0}/\Gamma_{a}$ \cite{Cecil1992}                    & 37(6) \\
      $\sigma_{p\gamma0}$ \cite{He2016} + $\Gamma_{\gamma0}$ \cite{Friebel1978}                       & 38(6) \\
      \bottomrule
    \end{tabular*}
  \end{threeparttable}
\end{table}

All of the cross section measurements of the $(p,\alpha)$
channel were performed with a small energy sensitive detector placed at various angles. The
measured energy spectrum was then extrapolated to 0 and integrated. \mrefs
\cite{Anderson1974,Davidson1979} performed a linear extrapolation, while \mref
\cite{Becker1987} used a sequential decay model. Although Becker \etal note their model
performed poorly at this resonance, it does \emph{not} explain the discrepancy between the
measurements. The key difference is the choice of normalization for the $\alpha_{1}$
measurement where Becker \etal argue that their detector has either detected the primary alpha
particle $\alpha_{1}$
or the two secondary $\alpha$
particles from the subsequent \Be break-up. Thus, they divide their count number by 2. On the
contrary, \mrefs \cite{Anderson1974,Davidson1979} argue they observe one out three final state
$\alpha$
particles and divide by a factor of 3.  The probability for detecting both secondary $\alpha$
particles in a single detector was discussed theoretically by Wheeler in 1941
\cite{Wheeler1941}. The probability depends on the opening angle between the secondary $\alpha$
particles and the aperture of the detector. The opening angle in turn depends on the energy of
the \Be system with respect to the $\alpha$
threshold.  This is small for the \Be GS but quite significant for the first excited
state. Based on information provided in \mref \cite{Becker1987} we have estimated their maximum
detector aperture to be of the order of $\SI{3}{\degree}$.
In this case the probability for a double hit is minuscule -- even for the $\alpha_{0}$
channel. Thus, the $\alpha_{1}$
results by Becker \etal should most likely be scaled by a factor of $2/3$
making it consistent with the other two measurements.  The astrophysical NACRE evaluation
\cite{Angulo1999} explicitly mentions this discussion in their $\Boron(p,3\alpha)$
evaluation, where they use the dataset of Becker \etal for its recommended value while using
the dataset of \mrefs \cite{Davidson1979,Segel1965} as a lower limit. The updated evaluation
NACRE II \cite{Xu2013} is less cautious and relies solely on Becker \etal

The magnitude of the $\sigma_{p\alpha}$ cross section has implications
beyond nuclear structure. For example the $\Boron(p,3\alpha)$
reaction is a candidate for a fusion reaction generating energy
without neutrons in the final state, see \textit{e.g.} \cite{Moreau1977}. The
rate of this reaction at the energies relevant for a fusion
reactor is mainly determined by the \SI{16.1}{MeV} $2^+$ and the higher lying
\SI{16.6}{MeV} $2^-$ resonances. The proton width is related to the
$\Boron(d,p)\BoronT$ reaction by isospin symmetry. In turn, that
reaction is used to deduce the $\Boron(n,\gamma)\BoronT$ reaction cross
section, which may play a role in the astrophysical r-process \cite{Lee2010}.

The object of this paper is to remeasure $\sigma_{p\alpha}$ in order
to clarify the situation. The measurement will circumvent the
normalization ambiguity by observing all three particles in
coincidence using an array of large area segmented silicon
detectors. In this paper we will only address the cross section of the
\SI{16.1}{\MeV} state, but the discussion on normalization applies
universally to all measurements of this reaction, where the cross
section is inferred from a single detector.

\section{Experiment}
\label{sec:experiment}

\begin{figure}[t]
  \includegraphics[width=\columnwidth]{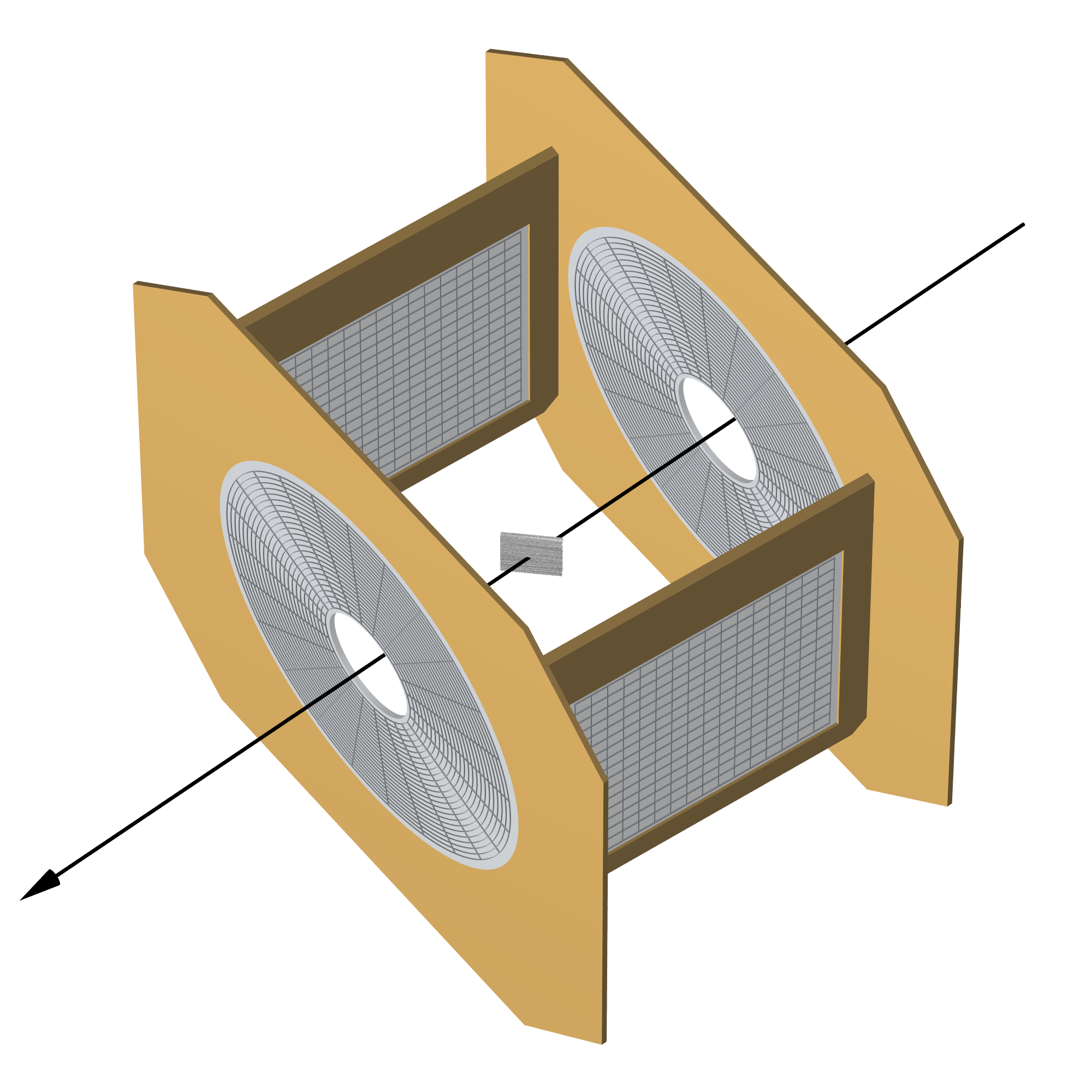}
  \caption{Schematic drawing of the experimental setup.
    An arrow indicates the incoming proton beam.
    The enriched boron target was oriented \SI{45}{\degree} with respect to the beam axis.
    Two quadratic and two annular double sided silicon strip detector were used to detect
    outgoing particles.
    Front and back segmentation is shown simultaneously for clarity. 
  }
  \label{fig:setup}       
\end{figure}

A thin foil of \Boron, oriented \SI{45}{\degree} with respect to the beam axis, was bombarded
with a beam of \ce{H_3^{+}} molecules. A resonance scan was conducted between 460 and
\SI{600}{\keV} and afterwards data was collected for 30 hours at \SI{525}{\keV}.  The beam was
provided by the \SI{5}{\mega V} Van de Graaff accelerator at Aarhus University and the beam
spot was defined by a pair of $1\times1$
\si{mm} vertical and horizontal slits. The energy of the accelerator was prior to the
experiment energy calibrated using narrow $(p,\alpha)$ resonances
in \ce{^{27}Al} and the energy resolution was found to be better than a few keV for single charged
beams. It should be noted that the energy stability is trifold improved for \ce{H_3^{+}}.

Upon impact with the target foil the \ce{H_3^{+}} molecules will break up and additional
electron stripping, neutralization and scattering might occur. This affected the integrated beam
current, which was measured with a Faraday cup \SI{1}{\m} downstream of the target. The combined
effect of this can be determined from the ratio of the observed current both with and without a
target foil in the beam. At the beginning of the experiment the effective charge state was
determined to be $\num{1.72(5)}e$. However, this ratio was observed to change during the
experiment. We attribute this to carbon build-up on the target. The change in charge state was
$\SI{4.33(4)E-3}{\micro C^{-1}}$. Correcting for this, a total of \SI{61(6)}{\micro \coulomb}
was collected on the resonance. 

The target was produced at Aarhus University by evaporation of \SI{99}{\%} enriched \Boron onto
a \SI{4}{\micro g \per \cm\squared} carbon backing. The thickness was measured by bombarding
the target with \SI{2}{\MeV} $\alpha$ particles and the boron layer either facing towards or away
from the beam. Assuming a two layer target the boron thickness can be inferred from the energy
shift of particles scattered off the carbon layer using the procedure of
\mref \cite{Chiari2001}, but including a correction for the changed stopping power of the
scattered particle.
\begin{equation}
  \label{eq:thickness}
  t = \frac{\delta E}{K(\theta)S(E_{b}) + S(K(\theta)E_{b})/\cos\theta},
\end{equation}
where
$E_{b}$
the beam energy, $S$
the stopping power,
$\delta E$
the energy difference and
$K$
is the kinematic factor for the laboraty scattering angle, $\theta$, defined as
\begin{equation}
  K = \frac{m_{b}\cos \theta + \sqrt{m_{t}^{2}-m_{b}^{2}\sin^{2} \theta}}{m_{b} + m_{t}},
\end{equation}
where $m_{t}, m_{b}$ is the mass of the target and beam ion respectively.
The cosine factor in \meqref{eq:thickness} corrects for the increased path length for the
scattered particle.  The result is \SI{39(3)}{\micro g \per \cm\squared}.  The energy loss for
a \SI{525/3}{\keV} proton through this target is \SI{23(2)}{\keV} according to the SRIM
stopping power tables \cite{SRIM}.

Charged particles were observed with an array of double sided silicon strip detectors (DSSD)
giving a simultaneous measurement of position and energy. A sketch of the array can be seen on
Figure~\ref{fig:setup}. The array consisted of two annular DSSD (S3 from Micron Semiconductors)
placed \SI{36}{\mm} up- and downstream of the target; covering the angles between 140 and
\SI{165}{\degree} and 23 and \SI{36}{\degree} respectively. Each annular ring is approximately
\SI{1}{\mm} wide with an approximate \SI{2}{\degree} resolution in polar angle. Additionally,
two quadratic DSSDs (W1 from Micron Semiconductors) were placed \SI{40}{\mm} from the target
center at an angle of \SI{90}{\degree} with respect to the beam axis. These covered angles
between \SI{60}{\degree} and \SI{120}{\degree}. All pixels are $3\times3$
mm with an approximate angular resolution of \SI{4}{\degree}.

\section{Analysis}
\label{sec:analysis}

The analysis is structured in the following way.  First a resonance scan is shown for the
$\alpha_{0}$
channel.  This is followed by an extraction of the $\alpha_{0}$
angular distribution and cross section from the multiplicity 1 data. Building upon this follows
the analysis of the triple events i.e. events with exactly three alpha particles and afterwards
a brief discussion of how the detection efficiencies for the $\alpha_{0}$
and $\alpha_{1}$ channel have been determined.

\begin{figure}[t]
  \includegraphics[width=\columnwidth]{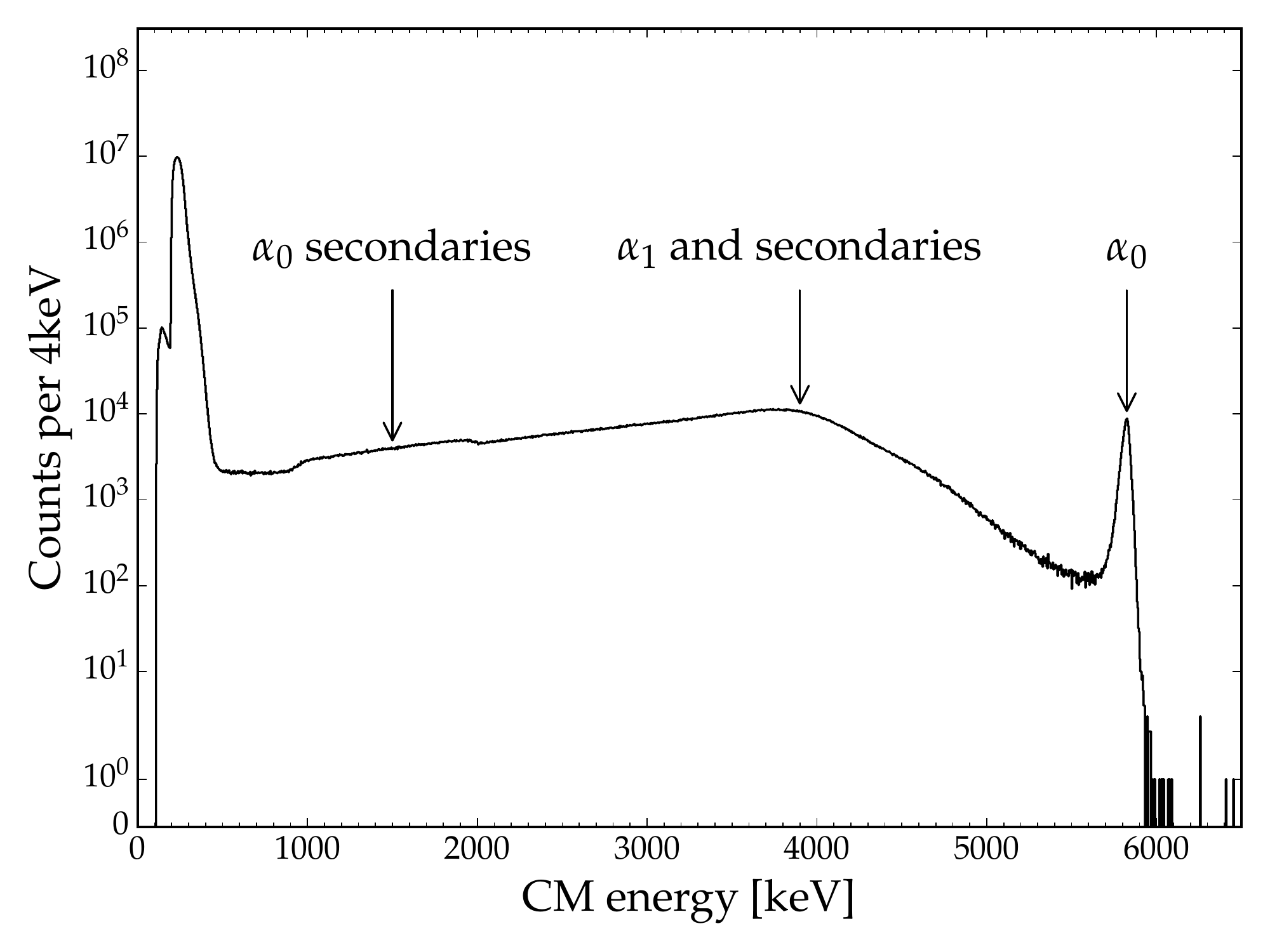}
  \caption{%
    Full CM energy spectrum without any cuts. The high energy peak corresponds to the primary
    $\alpha$ particle, $\alpha_{0}$.
  }
  \label{fig:singles}       
\end{figure}

\subsection{Singles analysis}
\label{sec:singles}

Assuming all ejectiles to be $\alpha$
particles the center-of-mass (CM) energy can be determined from the detected position and
energy. The full spectrum, without any cuts, is shown in \fref{fig:singles}.  The spectrum
shows a clear peak at \SI{5.8}{\MeV}, which is consistent with a sequential decay of the
\SI{16.1}{\MeV} state via the GS of \Be. The $\alpha$ particle giving rise to this peak will be
referred to as the primary $\alpha$ particle. Below the peak is a broad asymmetric distribution,
which consists of secondary $\alpha$ particles and $\alpha$ particles from the break-up via the first
excited state of \Be. At low energy the proton peak is
visible. It is double peaked since energy loss corrections are applied as if it was an $\alpha$
particle.

\begin{figure}[t]
  \includegraphics[width=\columnwidth]{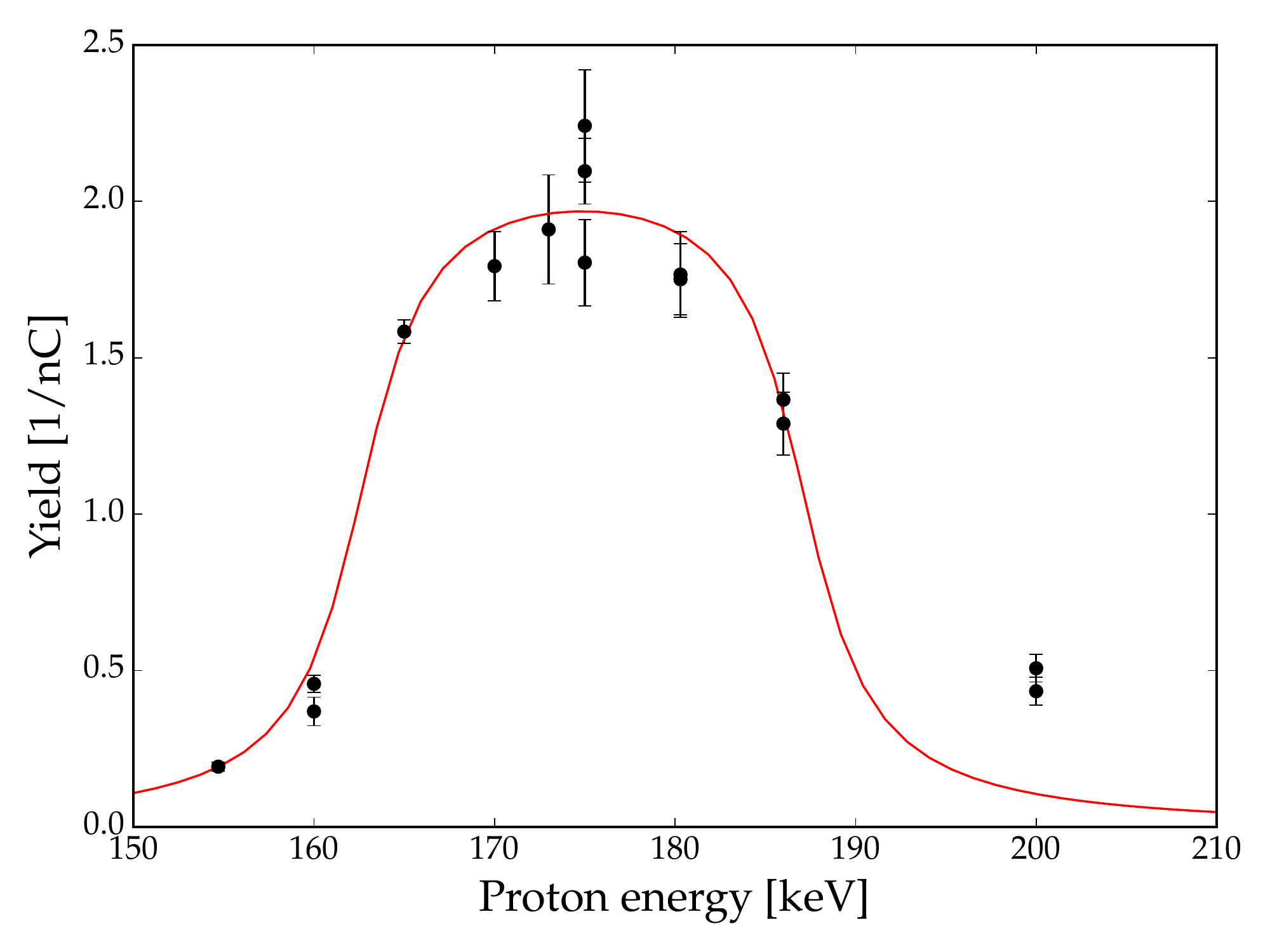}
  \caption{Scan of the \SI{162}{\keV} resonance. The individual data points corresponds to the
    $\alpha_{0}$ yield, while the solid curve is the best fit to \meqref{eq:yield}.}
  \label{fig:scan}       
\end{figure}
The primary $\alpha$ particle is selected by requiring $\ECM > \SI{5.65}{\MeV}$. \fref{fig:scan}
shows the $\alpha_{0}$ yield as a function of proton energy. The yield is clearly
resonant and peaks at $\unsim \SI{175}{\keV}$. The curve shown in the figure is the best fit to
the thick target yield for a Breit-Wigner shaped resonance~\cite{Fowler1948}
\begin{align}
  \label{eq:yield}
  Y = \Bigg[\tan^{-1} \frac{E_{p}-E_{r}}{\Glab/2} &- \tan^{-1} \frac{E_{p}-E_{r}-\Delta
                                                    E}{\Glab/2}\Bigg] \notag \\
                                               &\times \frac{\Glab\, \sigma_{\mathrm{BW}}(E=E_{r})}{2\epsilon} \eta,  
\end{align}
where $\Glab$
is the resonance width in the lab system,
$E_{p}$
is the beam energy, $E_{r}$
the resonance energy, $\Delta E$
the energy loss through the target, $\eta$ the detection efficiency, $\sigma_{\mathrm{BW}}$
the resonant Breit-Wigner cross section and $\epsilon = \frac{1}{N}\frac{dE}{dx}$,
where $N$ is the number density of target nuclei and $\frac{dE}{dx}$ the stopping power. Using
the factor outside the parenthesis as a arbitrary scaling factor, the best fit was achieved
with $\Delta E = \SI{24.5(9)}{\keV}$, $E_{r} = \SI{162.6(5)}{\keV}$ and $\Glab$ fixed to
$12/11\cdot\SI{5.28}{\keV}$. The target thickness is consistent with the result obtained from
$\alpha$-scattering and the resonance energy fits with the recommended literature value \cite{Ajzenberg-Selove1990}.

\begin{figure}[t]
  \includegraphics[width=\columnwidth]{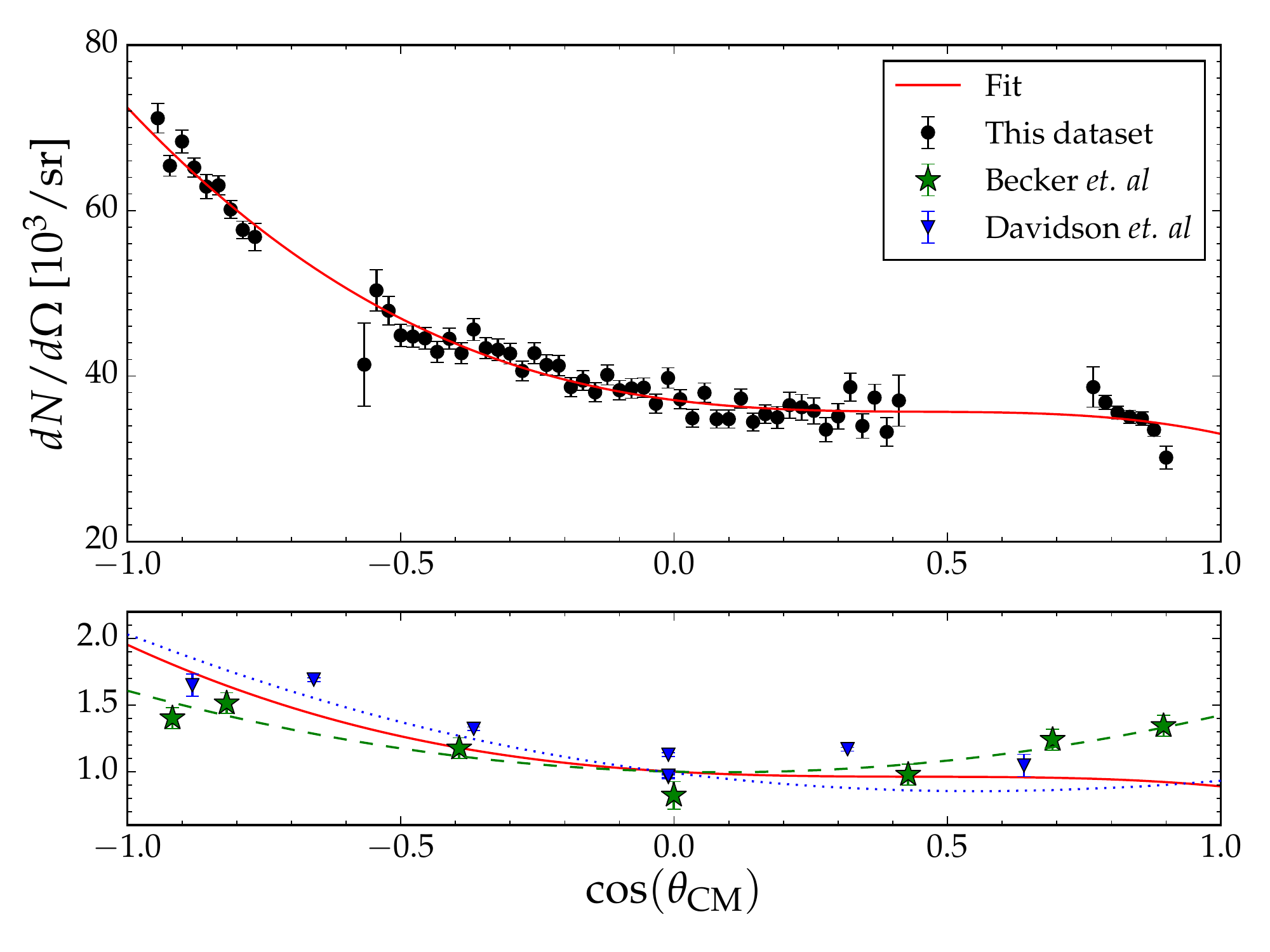}
  \caption{Angular distribution of this dataset corrected for solid angle along with the best
    fit to \meqref{eq:angular}. The datasets of \mref \cite{Becker1987,Davidson1979} have been
    rescaled to coincide with the fit at \SI{90}{\degree}.  }
  \label{fig:angular}       
\end{figure}

The $\alpha_{0}$
angular distribution relative to the beam axis was extracted for the long runs at
$E_{p} = \SI{175}{\keV}$.
The angular distribution, corrected for solid angle, can be seen in \fref{fig:angular}.
The solid line shows the
best fit to the lowest five Legendre polynomials.
\begin{equation}
  \label{eq:angular}
  W(\theta) = A\Bigg[1+\sum_{i=1}^{4}a_{i}P_{i}(\cos\theta)\Bigg].
\end{equation}
The coefficients providing the best fit are 
$a_{1} = \num{-0.358(7)}$,
$a_{2} = \num{0.249(2)}$, $a_{3} = \num{-0.106(11)}$, $a_{4} = \num{-9(20)E-3}$ and $A = \SI{4.253(17)E4}{sr^{-1}}$. 
The lower panel of the figure shows the fit rescaled, $W(\SI{90}{\degree}) = 1$,  along
with the data from \mrefs \cite{Becker1987,Davidson1979}, which have been rescaled to coincide
with the fit at \SI{90}{\degree}. Good agreement is observed with \mref \cite{Davidson1979}
while the symmetric behavior seen by \mref \cite{Becker1987} cannot be reproduced. 

The total number of counts is found by integration of \meqref{eq:angular} i.e. $4\pi A$.
This can be related to the resonant Breit-Wigner cross section using \meqref{eq:yield}
\begin{equation}
  \label{eq:single-cross}
  \sigma_{p\alpha,0} = \SI{2.1(2)}{\mb}.
\end{equation}
The main uncertainty is the variation in the effective charge state.

\subsection{Extraction of triple events}
\label{sec:triples}

The end goal of this analysis step is to extract tuples of particles consistent with a decay of
the \SI{16.1}{\MeV} state in \Carbon into three $\alpha$ particles. It applies the methods
described in \mref \cite{Alcorta2009}.

\begin{figure}[t]
  \includegraphics[width=\columnwidth]{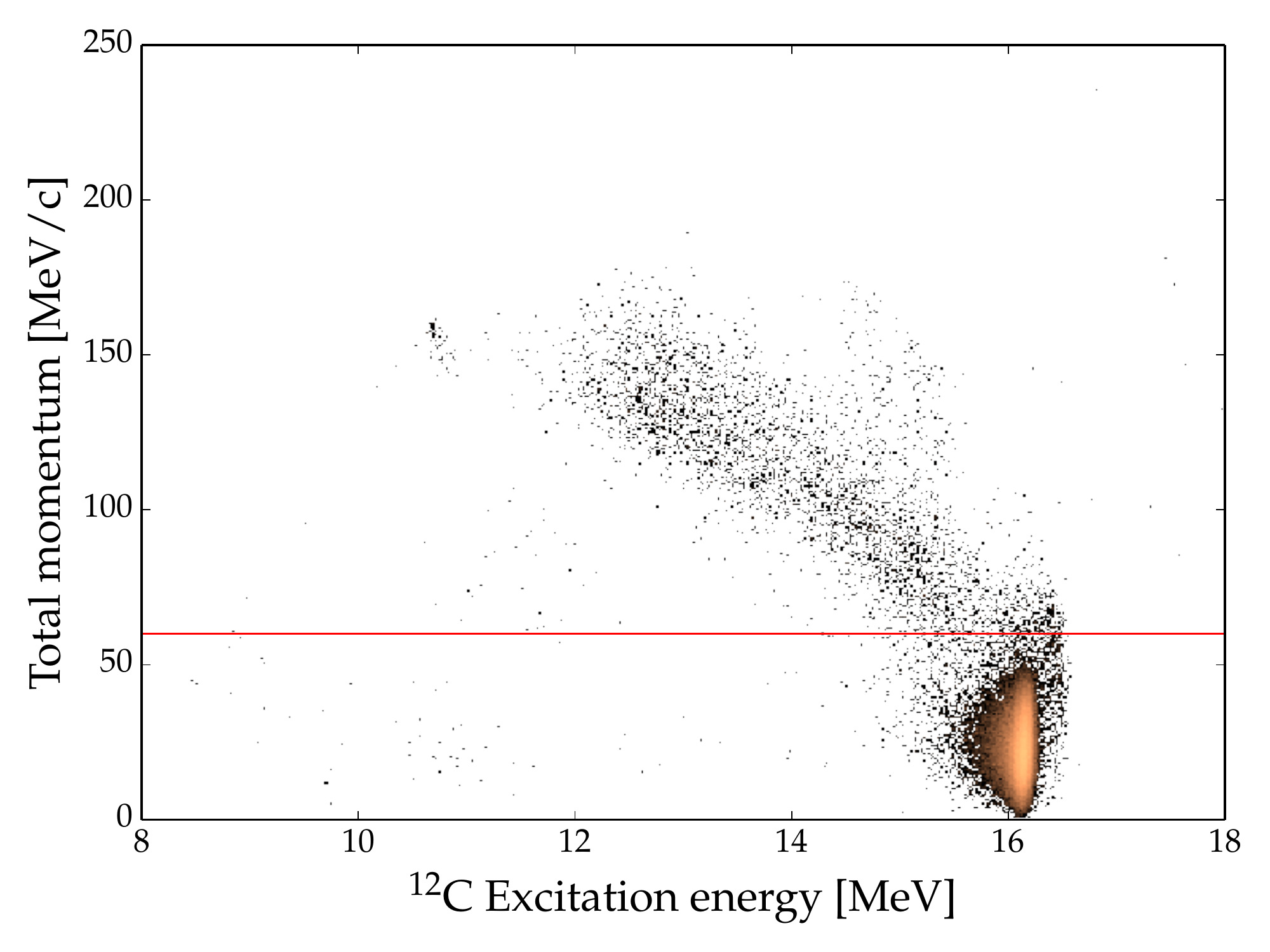}
  \caption{Total momentum of the three particles in the CM vs. the calculated excitation energy
    of \Carbon. The red line corresponds to the cut placed at \SI{60}{\MeV\per c}.}
  \label{fig:pExC12}       
\end{figure}

The first and simplest requirement is that at least three particles must be detected in an
event. This massively reduces the data, since the majority of events consist of elastically
scattered protons. Furthermore, it is required that all three particles are detected within
\SI{30}{ns} of each other. This reduces the background from random
coincidences with protons significantly while $> \SI{99}{\%}$ of good events survive. Additionally, it is required that the sum of CM angles between the
CM position vectors must be larger than \SI{350}{\degree}. All particles surviving these cuts are
assumed to be $\alpha$
particles. From the detected energy and position it is possible to calculate the four momentum
of each particle. From these, one can calculate the total momentum in the CM and the \Carbon
excitation energy. This is shown in \fref{fig:pExC12}, which has a distinct peak at
\SI{16.1}{\MeV}. Interestingly, weak peaks are visible at low total momentum and excitation
energy lower than \SI{16.1}{\MeV}. These correspond to $\gamma$
transitions in \Carbon as observed in \mref \cite{Laursen2016b}. Projecting the individual
energy of the high momentum events it is clear that these correspond to events with one proton
and two $\alpha$ particles. Hence, all events with $p_{\cm} > \SI{60}{\MeV\per c}$ are removed.

The classification of whether a tuple corresponds to a decay via the GS or first excited state,
can be done based on whether the CM energy of the most energetic particle lies within the high
energy peak in \fref{fig:singles}. This is the same cut used in sect. \ref{sec:singles}. With this classification the count numbers for the two
channels are
\begin{align}
  \label{eq:count}
  N_{0} &= \num{3.318(18)E4}\\
  N_{1} &= \num{4.33(2)E4},
\end{align}
where the uncertainties are due to counting statistics. 

\subsection{Detection efficiency of the $\alpha_{0}$ channel}
\label{sec:effic-gs}

In order to relate the observed number of counts to a yield it is necessary to determine the
detection efficiency. For the ground state this is simple. A beam with a $1\times1$
\si{mm} profile was generated and propagated to a random depth in the target. Here $\alpha_{0}$
was generated and emitted according to the observed angular distribution in
\fref{fig:singles}. The secondary particles were ejected isotropically according to
conservation of angular momentum. These particles were propagated out of the target and into
the detectors. Energy loss was taken into account using the SRIM energy loss tables
\cite{SRIM}.  The output of the simulation had a structure which was identical to the data and
was thus subjected to the same analysis. From the survival ratio an acceptance was determined
\begin{equation}
  \label{eq:eff-gs}
  \eta_{0} = \SI{7.1(3)}{\%}.
\end{equation}
Correcting for the efficiency gives a cross section of
\begin{equation}
  \label{eq:cross-gs-3}
  \sigma_{p\alpha,0} = \SI{2.0(2)}{\milli \barn},
\end{equation}
which is consistent with the singles analysis.

\subsection{Detection efficiency of the $\alpha_{1}$ channel}
\label{sec:effic-ex}



The detection efficiency depends heavily on the \Be excitation energy as it determines the
opening angle between the secondary $\alpha$ particles. Laursen \etal found that their sequential
decay model fully described their data \cite{Laursen2016}. Thus, events were generated using
this model. Propagation and energy loss was done with the same procedure as described in the previous section. From the survival ratio
the acceptance was determined to be
\begin{equation}
  \label{eq:eff-ex}
  \eta_{1} = \SI{0.49(5)}{\%}.
\end{equation}
This yields a cross section to the excited channel of
\begin{equation}
  \label{eq:cross-first-ex}
  \sigma_{p\alpha,1} = \SI{38(5)}{\milli \barn}.
\end{equation}

\section{Discussion}
\label{sec:discussion}

Both values determined for the $\alpha_{0}$
cross section are consistent with the measurement by Becker \etal The weighted
cross section is
\begin{equation}
  \label{eq:ca0-res}
  \sigma_{p\alpha,0} = \SI{2.03(14)}{\mb}.
\end{equation}
Comparing the angular distribution in \fref{fig:angular} with previous measurements, good
agreement is observed for the region around \SI{90}{\deg}. However, while
\mref \cite{Becker1987} finds the distribution to be nearly symmetric around \SI{90}{\deg}, this
conclusion is not supported by the present measurement or
\mref \cite{Davidson1979}. Importantly, the integrated cross section
is not very sensitive to the large angle behavior, which explains the good agreement obtained
nevertheless.

\begin{figure}[t]
  \includegraphics[width=\columnwidth]{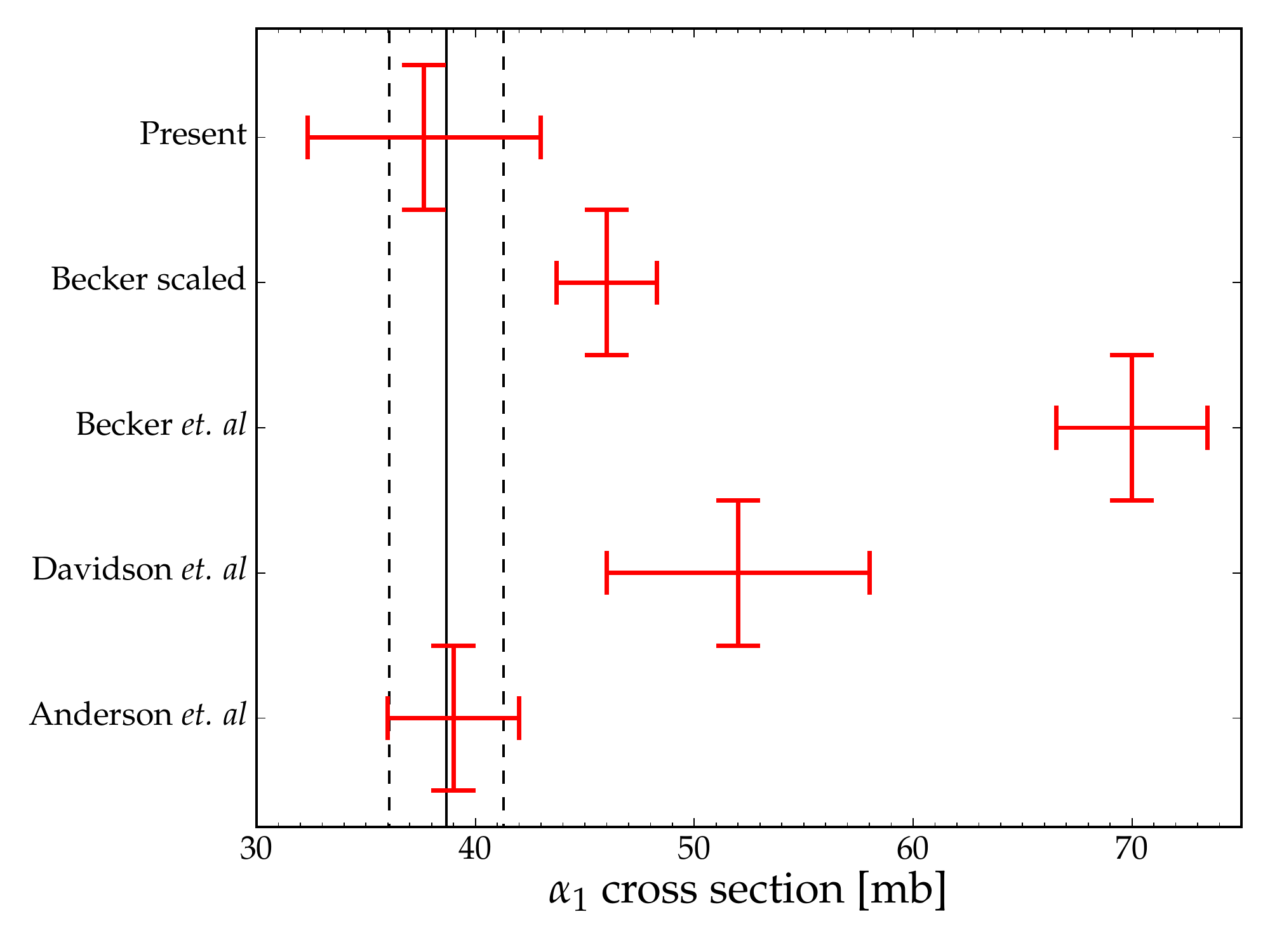}
  \caption{Comparison of the present $\alpha_{1}$
    cross section with the measurement of
    Refs. \cite{Anderson1974,Davidson1979,Becker1987}. Becker scaled is the data from
    Ref. \cite{Becker1987} scaled by $2/3$. The full line is the mean recommended value while
    the dashed lines show the one sigma limit. See text for details. 
  }
  \label{fig:compare}       
\end{figure}

In order to compare $\sigma_{p\alpha,{1}}$
between the different measurements, it is computed as
$\sigma_{p\alpha,{1}} = \sigma_{p\alpha} - \sigma_{p\alpha_{0}}$
for \mrefs~\cite{Anderson1974,Davidson1979}. The result is shown in \fref{fig:compare} along
with the present $\alpha_{1}$
cross section and that of \mref \cite{Becker1987}. From the figure the excellent argeement
between the present measurement and \mref \cite{Anderson1974} can be observed. Both
values deviate more than $2\alpha$
from the measurement of \mref \cite{Davidson1979}. We suspect this is due to an
overall normalization problem in \mref \cite{Davidson1979}. The original measurement by
\mref \cite{Becker1987} differs significantly from all other measurements, but if rescaled by a
factor $2/3$,
corresponding to the different normalization choice, the data point is in agreement within
the errors. However,
in light of the systematic problems
reported by \mref \cite{Becker1987} for the \SI{16.1}{\MeV}
resonance the value is not included in the recommended value%
\footnote{%
  The short-comings of their model at the \SI{16.1}{\MeV} resonance
  is due to them neglecting interference terms by summing incoherently over different $\alpha$
  permutations. The importance of interference was demonstrated in the work
  of \mrefs \cite{Phillips1965,Schafer1970}.
}%
. Instead the recommended
$\alpha_{1}$
cross section is based on the results from the present experiment and \mref~\cite{Anderson1974}
\begin{equation}
  \label{eq:ca0-res}
  \sigma_{p\alpha,1} = \SI{39(3)}{\milli \barn}.
\end{equation}
Similarly the recommended total $\alpha$ cross section is
\begin{equation}
  \label{eq:cross-tot}
  \sigma_{p\alpha} = \SI{41(3)}{\milli \barn}.
\end{equation}

The the ratio of the two $\alpha$ channels from the present measurement is \num{19(3)}, which is
consistent with both previous measurements. Combining all three measurements yields
\begin{equation}
  \label{eq:4}
  \frac{\sigma_{p\alpha,1}}{\sigma_{p\alpha,0}} = \num{19.9(14)}.
\end{equation}

\section{Derived partial widths}
\label{sec:derived-widths}

Using the present measurement of the $(p,\alpha)$ cross section the partial proton width can be
determined using \meqref{eq:width}
\begin{equation}
  \label{eq:this-proton-width}
  \Gamma_{p} = \SI{19(3)}{\eV},
\end{equation}
while using the combined cross section yields
\begin{equation}
  \label{eq:wa-proton-width}
  \Gamma_{p} = \SI{19.7(13)}{\eV}.
\end{equation}
Both values are consistent, within the errors, with that of the latest compilation
\cite{Kelley2017}, but not with the value favored in the recent review in \mref
\cite{Laursen2016b}.

Combining this proton width with the results of He \etal and $\Gamma$, the
partial gamma widths can be determined
\begin{align}
  \Gamma_{\gamma0} &= \SI{0.66(9)}{\eV} \\
  \Gamma_{\gamma1} &= \SI{18(2)}{\eV}.
\end{align}
These values are consistent with the latest compilation \cite{Kelley2017}, but
inconsistent with a direct measurement using inelastically scattered electrons, which measured
$\Gamma_{\gamma0} = \SI{0.346(41)}{\eV}$
\cite{Friebel1978}. While direct measurements should generally be favored, in this case there
is multiple independent and consistent measurements of the remaining parameters. In order to
resolve this discrepancy we suggest that $\Gamma_{\gamma0}$ is remeasured in a direct manner. The
argument in the recent review hinged on this value being correct \cite{Laursen2016b}. 

Combining the improved total $\alpha$
cross section with the $\gamma$ cross sections measured by He \etal the branching ratio can be determined
\begin{align}
  \label{eq:cecil-ratio}
  \frac{\Gamma_{\gamma0}}{\Gamma_{\alpha}} &= \num{1.18(13)E-5} \\
  \frac{\Gamma_{\gamma1}}{\Gamma_{\alpha}} &= \num{3.2(3)E-3}, 
\end{align}
which is inconsistent with the measurement by Cecil \etal Considering the general
spread of the measured $\alpha$ cross sections, the most likely explanation for this discrepancy is
that Cecil \etal have overestimated the $\alpha$ yield.

Using the updated proton width the ratio between the reduced proton and neutron width can be
computed. The analysis in \mref \cite{Monahan1971} was performed with $\Gamma_{p} = \SI{69}{\eV}$ and
an updated value can be computed by scaling accordingly
\begin{equation}
  \frac{\gamma_{n}^{2}}{2\gamma_{p}^{2}} = 0.63.
\end{equation}
This shows a similar degree of isopin
symmetry as the other bound states analysed in the $A=12$ system \cite{Monahan1971}.
\section{Conclusion}
\label{sec:conclusion}

Using the $p + \Boron$
reaction, the break-up of the \SI{16.1}{\MeV} state in \Carbon into three $\alpha$ particles has
been studied using an array of large area segmented silicon detectors in close geometry. The
decay via the ground state of \Be has been studied both with detection of single particles and
coincident detection of all three $\alpha$ particles. The derived cross sections are internally
consistent and the combined result is
\begin{equation}
  \sigma_{p\alpha,0} = \SI{2.03(14)}{\mb}, 
\end{equation}
which is consistent with the result of \mref \cite{Becker1987}.

Currently, there exists multiple incompatible measurements of the decay via the first excited
state of \Be. This channel was studied using coincident detection of all three $\alpha$ particles.
The coincidence acceptance was determined using the decay model of \mref \cite{Laursen2016}
yielding a model dependent cross section
\begin{equation}
  \sigma_{p\alpha,1} = \SI{38(5)}{\milli \barn}. 
\end{equation}
which is, within the errors, consistent with \mref \cite{Anderson1974} but not \mrefs
\cite{Davidson1979,Becker1987}.

The inconsistency with \mref \cite{Becker1987} is due to their claim of having a substantial
chance of detecting two out of three particles with a single detector. This was discussed based
on \mref \cite{Wheeler1941}. The chance of this is minuscule and hence the entire $\alpha_{1}$
dataset of \mref \cite{Becker1987} should be rescaled by a factor $2/3$.
This has a significant impact on the recommended astrophysical reaction rate, as both NACRE
\cite{Angulo1999} and NACRE II \cite{Xu2013} have based their recommended values on the dataset
provided by \mref \cite{Becker1987}. The recommended reaction rate should thus be scaled
accordingly. 
In addition, this also has implacations for the expected yield from an aneutronic fusion reactor.

Combining the present measurement of $\sigma_{p\alpha}$
with that of \mref \cite{Anderson1974} a refined partial proton width of
$\Gamma_{p} = \SI{19.7(13)}{\eV}$
was deduced.  This in turn, was used to determine the partial gamma widths
$\Gamma_{\gamma0} = \SI{0.68(8)}{\eV}$
and $\Gamma_{\gamma1} = \SI{18(2)}{\eV}$,
using the combined $\gamma$
cross sections reported by \mref \cite{He2016}. The value for $\Gamma_{\gamma0}$
differs by roughly a factor of 2 from the direct measurement of \mref \cite{Friebel1978}. Hence,
we recommend that $\Gamma_{\gamma0}$
is remeasured. Based on these results, we can no longer recommend the proton width deduced in
\mref \cite{Laursen2016b}. 

Additionally, improved $\gamma$-$\alpha$
branching ratios are derived. These are roughly a factor of 2 larger than the measurements
published by \mref \cite{Cecil1992}. We speculate that this discrepancy is most likely due to
\mref \cite{Cecil1992} having overestimated the $\alpha$ yield.

The recommended value for the proton width can be compared to the
spectroscopic factor for the analog state in $^{12}$B. 
By using the analysis presented in
\cite{Monahan1971} the ratio of the corresponding reduced widths is
0.63, which shows a similar degree of isopin
symmetry as the other bound states analysed in the $A=12$ system.
A modern calculation of the proton width would be highly
interesting.

\section{Acknowledgement}
\label{sec:acknowledgement}

We would like to thank Folmer Lyckegaard for manufacturing the target. We also acknowledge
financial support from the European Research Council under ERC starting grant LOBENA,
No. 307447.

\bibliography{IFA006-paper}

\end{document}